\documentclass[12pt,a4paper,english]{article}
\usepackage{float}
\usepackage{graphicx}
\usepackage{amsmath}
\usepackage{amsfonts}
\usepackage{latexsym}

\usepackage[dvips]{hyperref}

\newcommand{\noi}{\noindent}
\newcommand{\beq}{\begin{equation}}
\newcommand{\eeq}{\end{equation}}

\newcommand{\intd}{\mathrm{d}}

\newcommand{\bra}[1]{\langle #1 |}
\newcommand{\ket}[1]{|#1 \rangle}
\newcommand{\braket}[2]{\langle #1 | #2 \rangle}
\newcommand{\brakett}[1]{\langle #1 \rangle}
\newcommand{\brakettt}[3]{\langle #1 | #2 |#3 \rangle}

\usepackage{babel}

\begin{document}

\vspace*{-2cm}
\begin{flushright}
SISSA 84/2008/EP
\end{flushright}
\vspace{1cm}

\begin{center}
{\LARGE
Proposal to improve the behaviour of self-energy\\[2mm]
contributions to the S-matrix}
\vspace{1.5cm}

{\large
G\'{a}bor Zsolt T\'{o}th
\vspace{1cm}
}

\textit{
International School for Advanced Studies (SISSA),\\
 Via Beirut 2-4, 34014 Trieste,
Italy\\[5mm]
INFN, Sezione di Trieste, Italy\\[5mm]
Research Institute for Particle and Nuclear Physics, \\
Hungarian Academy of
Sciences, Pf.\ 49, 1525 Budapest, Hungary}\\[8mm]
e-mail:\ \  tgzs@cs.elte.hu

\end{center}
\vspace{1.5cm}

\begin{center}
{\bf Abstract}\\[8mm]

\begin{tabular}{p{14cm}}
{\small
A simple modification of the definition of the S-matrix is proposed.
It is expected that the divergences related to nonzero self-energies are considerably milder with the modified definition than 
with the usual one.
This conjecture is verified in a few examples using perturbation theory.
The proposed formula is written in terms of the total Hamiltonian operator and a free Hamiltonian operator and is therefore applicable in any case when these Hamiltonian operators are known.
}
\end{tabular}
\end{center}

\thispagestyle{empty}

\newpage

%\tableofcontents

%\newpage

\section{Introduction}

The S-matrix, introduced in \cite{H,JAW}, is a central object in
scattering theory and its calculation is of basic importance both in quantum
mechanics and in quantum field theory. There are certain well known problems that one encounters in scattering calculations;
one of them is that of the divergences caused by nonvanishing self-energies, which often appear in quantum field theory. In the present paper we concern ourselves 
mainly with this problem.

A general definition of the S-matrix (which will be reviewed briefly in section \ref{sec.rev}) can be given in
terms of the total
Hamiltonian operator $H$ of the interacting physical system
and a free
Hamiltonian operator $H_0$.
In quantum mechanics this definition can usually be applied without difficulty.
In quantum field theory, however, one encounters divergences in perturbation theory arising from 
disconnected vacuum-vacuum diagrams
and from radiative corrections on the external lines,
if one takes in a straightforward manner the quadratic part of $H$ as $H_0$ (see e.g.\ \cite{PS} chapter 4.\ for a detailed explanation). The disconnected vacuum-vacuum diagrams
and the radiative corrections on the external lines are known to correspond to
shifts of the vacuum energy and of the masses of the particles (these shifts are
called vacuum self-energy and self-masses, respectively) caused by the interaction part of $H$.

The mentioned divergences are distinct from the usual ultraviolet or infrared divergences and are present whenever the vacuum self-energy and the self-masses are nonzero. They are not specific to quantum field theory either; 
similar divergences appear
in any theory with nonzero self-energies, i.e.\ in any theory in which the eigenvalues of $H$ are shifted with respect to those of $H_0$.

This  problem of divergences is usually solved in the framework of renormalization theory.
If  $H_0$ and $H$ are chosen appropriately, then the disconnected vacuum-vacuum diagrams cancel out entirely and the radiative corrections on the external lines get replaced by factors called field strength renormalization constants.

In the present paper we propose a simple modification of the standard
definition of the \textit{in} and \textit{out} states and thus of the S-matrix.
The  modification that we propose consists in including certain phase factors in the
definition of the \textit{in} and \textit{out} states. Such a modification is allowed by the fact that physical states correspond to rays rather than to vectors, i.e.\ the phase of a state vector is not determined by the physical state that it represents.

We suggest that with our definition the complications related to self-energy corrections are considerably milder than with the usual definition, and that our definition allows more general pairs of $H_0$ and $H$ operators than the standard definition. These are the main virtues of our definition.

In section \ref{sec.rev} we recall briefly the standard formalism that we propose to modify.
In the subsequent section \ref{sec.main} we present our modified formalism, and in section \ref{sec.disc}
we discuss some of its features; in particular its relation to a well known formula proposed by Gell-Mann and Low \cite{1,FW}. The formulas in section \ref{sec.main} are quite general; they can be applied in a wide range of fields of physics, regardless of symmetries and other particular properties.
In section \ref{sec.example} we discuss various examples of different nature from quantum mechanics and from quantum field theory.
These examples are aimed to
provide some illustration for our definition and to demonstrate the cancellation of the divergences related to nonzero self-energies.
We also verify that the results yielded by our formalism agree with the results that can be obtained by means of the standard methods. For our calculations we use perturbation theory, which is described in appendix \ref{sec.pert}.
In appendix \ref{sec.reg} we
outline a regularization procedure for handling distributions (i.e.\ generalized functions) that occur in the calculations.
Finally, we mention that in accordance with the nature of the subject of this paper we decided to cite mostly books and only a few papers.

\section{The general formalism of scattering theory}
\label{sec.smatrix}

\subsection{A brief review of the standard definition of the S-matrix}
\label{sec.rev}

The S-matrix elements can be defined as
\begin{equation}
\label{eq.gdf}
S_{wv}=\braket{w,out}{v,in}
\end{equation} 
where $\ket{v,in}$ and $\ket{w,out}$ are suitable states called \textit{in} and
\textit{out}
states.

The  \textit{in} and \textit{out} states are defined in terms of two self-adjoint
operators $H$ and $H_0$, $H$ being the total Hamiltonian operator
describing the scattering problem and $H_0$ a free or reference
Hamiltonian operator. The definition involves eigenvectors of $H_0$ and  the time evolution operator
\begin{equation}
\label{eq.u2}
U_\epsilon(t_2,t_1)
=T\exp\left[-\frac{i}{\hbar}\int_{t_1}^{t_2}H_{I,\epsilon}(t)\, dt\right],
\end{equation}
where
\begin{equation}
\label{eq.pt1}
H_{I,\epsilon}(t)=e^{-\epsilon |t|} e^{\frac{i}{\hbar}H_0t}H_I e^{-\frac{i}{\hbar}H_0t}
\end{equation}
and
\begin{equation}
\label{eq.pt2}
H_I  =  H-H_0,
\end{equation}
and  the $T$ in (\ref{eq.u2}) denotes the time ordering:\\
$T[A_1(t_1)A_2(t_2)\dots A_n(t_n)]=
A_{k_1}(t_{k_1})A_{k_2}(t_{k_2})\dots A_{k_n}(t_{k_n})$, where $k_1,k_2,\dots,
k_n$ are determined by the condition $t_{k_1}>t_{k_2}>\dots >t_{k_n}$.
 It should be noted that
$U_\epsilon(t_2,t_1)$ is unitary and it has the
 properties
$U_\epsilon(t,t)=I$ (where $I$ is the identity operator),
$U_\epsilon(t_2,t_1)^{-1}=U_\epsilon(t_1,t_2)$
 and  $U_\epsilon(t_3,t_2)U_\epsilon(t_2,t_1)=
U_\epsilon(t_3,t_1)$.
An adiabatic switching is included in the above definition of $U_\epsilon(t_2,t_1)$, which serves 
as a regularization for the definition of the limits  $t_2\to \infty$ or $t_1\to -\infty$.
$\epsilon>0$ is the parameter of the adiabatic switching;
\begin{equation}
 \lim_{\epsilon\to 0}  U_\epsilon(t_2,t_1)=
U(t_2,t_1)=e^{\frac{i}{\hbar}H_0t_2}e^{-\frac{i}{\hbar}H(t_2-t_1)}e^{-\frac{i}{\hbar}H_0t_1}.
\end{equation} 
The eigenvalue for an eigenvector
$\ket{v}$ of $H_0$ will be denoted by $E_v^0$:
\begin{equation}
H_0\ket{v}  =E_v^0\ket{v}.
\end{equation}
The \textit{in} and \textit{out} states corresponding to an eigenvector $\ket{v}$ of $H_0$ are defined as
\begin{eqnarray}
\label{eq.G0in}
\ket{v,in} & = & \lim_{\epsilon\to 0^+} \lim_{T\to\infty} 
U_\epsilon(0,-T)\ket{v}\\
\label{eq.G0out}
\ket{v,out} &  = & \lim_{\epsilon\to 0^+} \lim_{T\to\infty} 
U_\epsilon(0,T)\ket{v},
\end{eqnarray}
so the S-matrix elements are
\begin{equation}
\label{eq.qmsmat}
\braket{w,out}{v,in} =\lim_{\epsilon\to 0^+}
\lim_{T\to\infty}\brakettt{w}{U_{\epsilon}(T,-T)}{v}.
\end{equation}
It is required that $\ket{v,in}$ and $\ket{v,out}$ be eigenvectors of
$H$. 
We note that in this section and throughout the paper multiple limits are understood to be evaluated
 one by one from right to left. It is always the $\epsilon \to 0^+$ limit
 which is evaluated last.

In quantum mechanics the above definitions can usually be applied without difficulty.
In quantum field theory, however, one usually finds that the $\epsilon\to 0^+$ limits in 
(\ref{eq.G0in}), (\ref{eq.G0out}) and (\ref{eq.qmsmat})
do not exist if one
takes the quadratic part of $H$ as $H_0$. As we mentioned in the introduction, this divergence can be associated with 
a shift of the vacuum energy and of the particle masses caused by the higher degree terms in $H$.
A standard  method to deal with this difficulty is to modify (renormalize) the pair $H_0$, $H$ used to produce the
\textit{in} and \textit{out} states.

For more detail on scattering theory we refer the reader
to \cite{PS,W,BD1,BD2,Iagolnitzer,Taylor,Newton, GG, Pir1, Pir2, DMS2}.
We note that instead of the adiabatic switching, other regularization methods such as taking
the abelian limit are also used in the literature (see e.g.\ \cite{Taylor,Newton}).

\subsection{The modified formulas}
\label{sec.main}

In this section we present the modifications that we propose in the definitions (\ref{eq.G0in}), (\ref{eq.G0out})
of the  $\ket{v,in}$ and $\ket{v,out}$ states, and thereby in the formula (\ref{eq.qmsmat}) for the S-matrix
elements. We do not change (\ref{eq.gdf}).

For the \textit{in} and \textit{out} states we propose the following definitions:
\begin{eqnarray}
\label{eq.Gin}
\ket{v,in} & = & \lim_{\epsilon\to 0^+} \lim_{T\to\infty} \Pi_{v,\epsilon}(-T)
U_\epsilon(0,-T)\ket{v}\\
\label{eq.Gout}
\ket{v,out} & = &\lim_{\epsilon\to 0^+} \lim_{T\to\infty} \Pi_{v,\epsilon}(T)
U_\epsilon(0,T)\ket{v},
\end{eqnarray}
where $ \Pi_{v,\epsilon}(T)$ is a complex
number of absolute value $1$, i.e.\ a phase factor, given by the formula
\begin{equation}
\label{eq.phase1}
 \Pi_{v,\epsilon}(T) =
\frac{\sqrt{\brakettt{v}{U_\epsilon(T,0)}{v}\brakettt{v}{U_\epsilon(0,T)}{v}}}{
\brakettt{v}{U_\epsilon(0,T)}{v}}.
\end{equation}
$U_\epsilon(t_2,t_1)$ is given by the same formula (\ref{eq.u2}) as in section \ref{sec.rev}.
The S-matrix elements are then given by
\begin{equation}
\label{eq.smf}
S_{wv}=
\braket{w,out}{v,in}= \lim_{\epsilon\to 0^+} \lim_{T\to\infty}
 \Pi_{w,\epsilon}(T)^*    \Pi_{v,\epsilon}(-T)
\brakettt{w}{U_\epsilon(T,-T)}{v}.
\end{equation}
The states $\ket{v,in}$ and $\ket{v,out}$ should be eigenvectors of
$H$ with eigenvalues that we denote by $E_{v,in}$ and $E_{v,out}$. We do not
require
that $E_{v,in}=E_{v}^0$ and $E_{v,out}=E_{v}^0$.
The eigenvalues $E_{v,in}$ and $E_{v,out}$ can be obtained from the equations
\begin{eqnarray}
\label{eq.energy1}
\lim_{\epsilon\to 0^+} \lim_{T\to\infty}
\frac{\brakettt{v}{HU_\epsilon(0,-T)}{v}}{
\brakettt{v}{U_\epsilon(0,-T)}{v}} & = & E_{v,in}\\
\label{eq.energy2}
\lim_{\epsilon\to 0^+} \lim_{T\to\infty}
\frac{\brakettt{v}{HU_\epsilon(0,T)}{v}}{
\brakettt{v}{U_\epsilon(0,T)}{v}} & = & E_{v,out}\ .
\end{eqnarray}

The definitions (\ref{eq.Gin}) and (\ref{eq.Gout}) are understood in the sense that if one wants to calculate a matrix element, for example, then one should do the 
calculation using\\ $\lim_{T\to\infty}\Pi_{v,\epsilon}(-T)U_\epsilon(0,-T)\ket{v}$ and $\lim_{T\to\infty}\Pi_{v,\epsilon}(T)U_\epsilon(0,T)\ket{v}$ 
with $\epsilon\ne 0$ first, and take the $\epsilon\to 0^+$ limit afterwards.

Often one has an orthonormal basis
$\{\ket{v_i}\}$ consisting of eigenvectors of $H_0$, where $i$ labels the elements of the basis, and one is interested in the \textit{in} and \textit{out} states, S-matrix elements and energies corresponding to these basis vectors.
We use the notation $E_i^0$, $E_{i,in}$, $E_{i,out}$, $ \Pi_{i,\epsilon}(T)$, $S_{ij}$, etc.\ for this case.

\subsection{Discussion of the proposed formulas}
\label{sec.disc}

In this section we discuss the formulas proposed in section \ref{sec.main}. In
order to clearly identify the various remarks, we present them in the form of a
numbered list.

\vspace{2mm}
\noi
1.) We stress that we do not
require that the spectrum of $H_0$ should be the same, even partially, as the spectrum of $H$.
In particular, it is not required that the particle masses corresponding to
$H_0$ should be the same as the  particle masses corresponding to
$H$.
Energy corrections,  in particular the shift
of the vacuum energy and of the particle
masses can be calculated using (\ref{eq.energy1}) and  (\ref{eq.energy2}).
We expect that generally $E_{v,in}=E_{v,out}$.

We also stress that in general neither $\lim_{T\to\infty} \Pi_{v,\epsilon}(-T)$ nor $\lim_{T\to\infty}U_\epsilon(0,-T)\ket{v}$ 
in (\ref{eq.Gin}) 
is convergent in itself in the $\epsilon \to 0^+$ limit (and a similar statement can be made about (\ref{eq.Gout})). 

It is worth noting that the formulation in section \ref{sec.main} 
does not make use of symmetries, fields and the existence of a vacuum state.
The formulas themselves are meaningful even for operators $H_0$, $H$ which do not describe scattering.

\vspace{2mm}
\noi
2.) The phase
factors  $ \Pi_{v,\epsilon}(T)$ and $ \Pi_{v,\epsilon}(-T)$  cancel out
in the  square of the absolute value of the S-matrix element
$\braket{w,out}{v,in}$:
\begin{equation}
S_{wv}S_{wv}^*=
\braket{v,in}{w,out}\braket{w,out}{v,in}
=\lim_{\epsilon\to 0^+} \lim_{T\to\infty}
\brakettt{v}{U_\epsilon(-T,T)}{w}\brakettt{w}{U_\epsilon(T,-T)}{v}.
\end{equation}

\vspace{2mm}
\noi
3.) 
The  formulas (\ref{eq.Gin}), (\ref{eq.Gout}) and (\ref{eq.smf})  can be rewritten as
\begin{eqnarray}
\label{eq.G2in}
\ket{v,in} & = & \lim_{\epsilon\to 0^+} \lim_{T\to\infty} \sqrt{X_{v,\epsilon}(-T)}
\frac{U_\epsilon(0,-T)\ket{v}}{
\brakettt{v}{U_\epsilon(0,-T)}{v}}\\
\label{eq.G2out}
\ket{v,out} & = & \lim_{\epsilon\to 0^+} \lim_{T\to\infty} \sqrt{X_{v,\epsilon}(T)}
\frac{U_\epsilon(0,T)\ket{v}}{
\brakettt{v}{U_\epsilon(0,T)}{v}}
\end{eqnarray}
and
\begin{equation}
\label{eq.smf2}
S_{wv}=
\braket{w,out}{v,in}=
\lim_{\epsilon\to 0^+} \lim_{T\to\infty}
\sqrt{X_{w,\epsilon}(T)}\sqrt{X_{v,\epsilon}(-T)}
\frac{\brakettt{w}{U_\epsilon(T,-T)}{v}}{\brakettt{w}{U_\epsilon(T,0)}{w}\brakettt{v}{U_\epsilon(0,-T)}{v}},
\end{equation}
where
\begin{equation}
\label{eq.x1}
X_{v,\epsilon}(T)
=\brakettt{v}{U_\epsilon(T,0)}{v}\brakettt{v}{U_\epsilon(0,T)}{v}.
\end{equation}
The right hand sides of  (\ref{eq.Gin}) and (\ref{eq.Gout}) written in the above form (\ref{eq.G2in}), (\ref {eq.G2out}) are
very similar to the formula introduced by Gell-Mann and Low in \cite{1} (see also \cite{FW}), 
which serves to produce the vacuum state of $H$ from the vacuum state of $H_0$.

The right hand sides of (\ref{eq.G2in}) and (\ref{eq.G2out}) and Gell-Mann and Low's 
formula (as written in \cite{FW}) differ in the factors $\sqrt{X_{v,\epsilon}(-T)}$ and $\sqrt{X_{v,\epsilon}(T)}$. These factors are real and are needed in order to maintain the correct normalization of the \textit{in} and \textit{out} states, which is important for the unitarity of the S-matrix. Of course, the formulas (\ref{eq.G2in}) and 
(\ref{eq.G2out}) also differ from Gell-Mann and Low's formula in the type and range of eigenvectors of $H_0$ to which they are applied.

A main feature of Gell-Mann and Low's formula is that the phase of the vector that it produces from a vector $\ket{v}$ is fixed with respect to $\ket{v}$. This is true for our formulas (\ref{eq.G2in}) and (\ref {eq.G2out}) as well, since the factors $\sqrt{X_{v,\epsilon}(-T)}$ and $\sqrt{X_{v,\epsilon}(T)}$ are real.
Let us 
introduce $\ket{v_{\epsilon}(-T),in}$ and $\ket{v_{\epsilon}(T),out}$ as
\begin{eqnarray}
\label{eq.GTin}
\ket{v_{\epsilon}(-T),in} & = & \Pi_{v,\epsilon}(-T)
U_\epsilon(0,-T)\ket{v}\\
\label{eq.GTout}
\ket{v_{\epsilon}(T),out} & = & \Pi_{v,\epsilon}(T)
U_\epsilon(0,T)\ket{v}.
\end{eqnarray}
We have 
\begin{eqnarray}
\label{eq.pr1}
 \sqrt{X_{v,\epsilon}(-T)} & = &
\braket{v}{v_{\epsilon}(-T),in}
 \\
\label{eq.pr2}
\sqrt{X_{v,\epsilon}(T)}
& = & \braket{v}{v_{\epsilon}(T),out}.
\end{eqnarray}
Equation (\ref{eq.pr1}) shows that if $\braket{v}{v_{\epsilon}(-T),in}\ne 0$, then the overall phase of  $\ket{v_{\epsilon}(-T),in}$ is fixed in such a way that $\braket{v}{v_{\epsilon}(-T),in}$ is real. The
same can be said about (\ref{eq.pr2}) and  $\ket{v_{\epsilon}(T),out}$.

The above feature of Gell-Mann and Low's formula is achieved by the denominator
$\brakettt{v}{U_\epsilon(0,-T)}{v}$. If this denominator
is not included, then the phase of the produced vector is generally not convergent as $\epsilon\to 0$
(see \cite{FW}).

Gell-Mann and Low's formula 
is known to be applicable to discrete nondegenerate eigenstates other than the vacuum as well (see \cite{FW}),
and it is not necessary for its applicability that the eigenstate of $H_0$ to which it is applied have
the same eigenvalue as the produced eigenstate of $H$.

We can say that our proposal in this paper is to modify Gell-Mann and Low's formula (as written in \cite{FW}) by including the factors 
$\sqrt{X_{v,\epsilon}(-T)}$ and $\sqrt{X_{v,\epsilon}(T)}$, and to apply it in this form to any eigenstate 
(not just to discrete nondegenerate eigenstates) with
the purpose of producing general \textit{in} and
\textit{out} states, which can contain any number of particles.
As we mentioned above, the factors $\sqrt{X_{v,\epsilon}(-T)}$ and $\sqrt{X_{v,\epsilon}(T)}$ are needed to maintain the correct normalization of the 
\textit{in} and \textit{out} states. The possible problems that can arise from the nondegeneracy of the eigenvector from which we intend to produce  \textit{in} and \textit{out} states will be discussed briefly in section \ref{sec.example} in the fifth example.

Concerning the requirement that the \textit{in} and \textit{out} states should be eigenvectors of $H$, Gell-Mann and Low's formula is known (see \cite{1,FW}) to produce an eigenvector of $H$ if the $\epsilon\to 0^+$ limit that it involves exists, and it is natural to expect that this remains true if the formula is applied to eigenvectors $\ket{v}$ of $H_0$ which are not discrete and nondegenerate.

\vspace{2mm}
\noi
4.) 
Let us assume that $\ket{v}$ is a discrete nondegenerate eigenvector. In a perturbative framework $\ket{v,in}$ and $\ket{v,out}$ will also be discrete nondegenerate eigenvectors, and they will belong to the same one-dimensional eigenspace, i.e.\ $\ket{v,in}=\alpha\ket{v,out}$ with some complex number $\alpha$.
$\braket{v}{v,in}$ and $\braket{v}{v,out}$ are both positive real numbers (they can be assumed to be nonzero in a perturbative framework), hence it follows that 
$\ket{v,in}=\ket{v,out}$ and $S_{vv}=\braket{v}{v}$ (see also \cite{FW}). This result holds in particular for the vacuum in quantum field theory. We expect that it also holds for one-particle states in Poincare symmetric theories.

\vspace{2mm}
\noi
5.)
In the case of Poincare symmetric theories we also expect that for a multi-particle state $\ket{k_1,k_2,...,k_n}$ of $H_0$ the total energy of $\ket{k_1,k_2,...,k_n,in}$ and $\ket{k_1,k_2,...,k_n,out}$
has the properties that $E_{k_1,k_2,...,k_n,in}=E_{k_1,k_2,...,k_n,out}$ and
$E_{k_1,k_2,...,k_n,in}-E_{\Omega_0,in}=\sum_{i=1}^n (E_{k_i,in}-E_{\Omega_0,in})$, where $\Omega_0$ denotes the vacuum state. The latter additivity property is expected because
$\ket{k_1,k_2,...,k_n,in}$ and $\ket{k_1,k_2,...,k_n,out}$ should be multi-particle eigenstates of $H$.

\vspace{2mm}
\noi
6.) We expect that instead of the exponential switching function $e^{-\epsilon|t|}$ other
functions can be used as well. For example, a linear switching function could also be used.
The term adiabatic switching does not, of course, refer to any physically real switching of the interaction.
In this paper we shall not discuss the precise role of the adiabatic switching prescription.
A feature of this method that is worth noting is that the unitarity of the time evolution operator is preserved 
for all values of $\epsilon$.

\vspace{2mm}
\noi
7.) In quantum theory the vector representing a physical state is fixed up
to a phase factor only (in other words, physical states correspond to rays),
so  in general we can say that the phases of the \textit{in} and \textit{out} states are
not determined:
one could multiply the right hand sides of  (\ref{eq.Gin}) and (\ref{eq.Gout}) by
any phase factors $\pi_{v,in}$ and  $\pi_{v,out}$.

\vspace{2mm}
\noi
8.) Since the  \textit{in}  and  \textit{out} states are eigenstates of $H$,
\begin{equation}
S_{wv}=0
\end{equation}
holds
if  $E_{w,out}\ne E_{v,in}$. This property expresses the conservation of energy.

\vspace{2mm}
\noi
9.) In the case of some $H_0$ and $H$ it can happen that (\ref{eq.Gin}) and
 (\ref{eq.Gout}) do
not produce eigenvectors of $H$ for
certain eigenvectors of $H_0$.
One of the situations in which this can be expected to occur is when
$H_I$ makes some particles unstable.

\vspace{2mm}
\noi
10.) One encounters infrared and ultraviolet divergences in many theories.
In this paper we do not discuss these divergences;
they can be handled in the same way as usual.

\vspace{2mm}
\noi
11.) Our formalism allows more general pairs of $H_0$ and $H$ than the standard formalism (i.e.\ (\ref{eq.G0in}), 
(\ref{eq.G0out}), (\ref{eq.qmsmat})), nevertheless
the right choice of $H$, $H_0$ and $\{ \ket{v_i}\} $ for a given physical system should be subject to
consideration in general.  We do not discuss the physical interpretation of $H_0$ and $\{ \ket{v_i} \}$ in this paper;
we regard them as auxiliary quantities which are used to produce, by means of the formulas (\ref{eq.Gin}) and (\ref{eq.Gout}), suitable \textit{in}  and \textit{out} states.

\vspace{2mm}
\noi
12.) In certain problems, e.g.\ in rearrangement scattering, one can have two different free Hamiltonian operators $H_0^A$
and $H_0^B$; one for the incoming states and one for the outgoing states
(see e.g.\ \cite{Newton}). 
Even if $H_0^A=H_0^B=H_0$, it might
be necessary to take two different sets of basis vectors $\{\ket{v_i}\}$ and $\{\ket{w_j}\}$ to define the \textit{in} and \textit{out} states.
For simplicity, we restrict ourselves to the case when $H_0^A=H_0^B=H_0$ and a single set of eigenvectors is used to define both the  \textit{in} and the \textit{out} states.
The extension of the discussion to the general case is straightforward.

\vspace{2mm}
\noi
13.) By orthonormality of the eigenvectors $\ket{v_i}$ of $H_0$ we mean
\begin{equation}
\braket{v_i}{v_j}=\delta (i,j),
\end{equation}
where $\delta (i,j)$ is
a Dirac-delta; $\delta (i,j)$=0 if $i\ne j$ and
\begin{equation}
\int \intd j\, \delta(i,j) g(j) =g(i),\qquad
\int \intd i\, \delta(i,j) g(i) =g(j),
\end{equation}
where $g(i)$ is any test function.
The integration $\int \intd i$ over the index set is
understood in a general sense; it may include summation over discrete parts of
the spectrum, for example.  In accordance with this, the Dirac-delta
$\delta(i,j)$ is also understood in a generalized sense.
The name eigenvector is often reserved to proper eigenvectors with finite norm, and eigenvectors belonging
to continuous parts of the spectrum are called improper eigenvectors. In this paper we
use the term eigenvector for both proper and improper eigenvectors.

\vspace{2mm}
\noi
14.) $\braket{v_i}{v_i}$ is often not a finite number and
 $\braket{v_i}{v_j}$ has to be regarded as a distribution.
When one applies perturbation theory or other methods to calculate (\ref{eq.smf}) and the other quantities in section \ref{sec.main}, this leads to expressions which are not well defined, in principle,
 since  (\ref{eq.smf}) will involve products and quotients of distributions.
One way to overcome this difficulty
is to introduce a regularization of the eigenvectors  $\ket{v_i}$.
Such a regularization prescription is outlined in appendix \ref{sec.reg}.
In some cases, however, it is sufficient to manipulate with
 $\braket{v_i}{v_j}$ in a formal manner, and the regularization can be
 avoided.

\vspace{2mm}
\noi
15.) As is well known,
the nonzero S-matrix elements (and also the components of the \textit{in} and
\textit{out} states) are usually not finite numbers, but the integrals
$\int \intd i\, \intd j\, \phi(a,i)^*\phi(b,j)S_{ij}$, where $\phi(a,i)$ and
$\phi(b,j)$ are suitable functions, are finite, i.e.\ the S-matrix is a
distribution (and a similar statement can be made about the  \textit{in} and
\textit{out} states).
The mentioned integral
is equal to $\braket{\Phi_a}{\Phi_b}$, where $\ket{\Phi_a}$ and $\ket{\Phi_b}$
are wave packets of  \textit{in} and \textit{out} states:
$\ket{\Phi_a}=\int \intd i\, \phi(a,i)\ket{v_i,out}$,
$\ket{\Phi_b}=\int \intd j\, \phi(b,j)\ket{v_j,in}$.

\vspace{2mm}
\noi
16.) In section \ref{sec.main} the quantities called S-matrix elements are defined for a certain set of
\textit{in}  and \textit{out} states that correspond to eigenvectors of $H_0$.
However, a matrix that could be called S-matrix was not defined.
In order to be able to define a complete S-matrix we
assume that an orthonormal set of vectors $\{ \ket{v_i}\} $ is chosen; then the matrix constituted by the S-matrix elements
$S_{ij}$ can be called the S-matrix. The
S-matrix elements for arbitrary superpositions of the \textit{in}  and \textit{out}
states $\ket{v_i,in}$ and $\ket{v_i,out}$ are determined by linearity:
let $\ket{\Phi_a}$ and $\ket{\Phi_b}$ be
superpositions of  \textit{in} and \textit{out} states:
$\ket{\Phi_a}=\int \intd i\, \phi(a,i)\ket{v_i,out}$,
$\ket{\Phi_b}=\int \intd j\, \phi(b,j)\ket{v_j,in}$.
The S-matrix element $S_{ab}$ is then
$S_{ab}=\braket{\Phi_a}{\Phi_b}=\int \intd i\, \intd j\, \phi(a,i)^*\phi(b,j)S_{ij}$.
It is important to note here that the $\ket{v}$ appearing in (\ref{eq.Gin}) and
(\ref{eq.Gout}) should always be an eigenvector of $H_0$, and that (\ref{eq.Gin}) and
(\ref{eq.Gout}) are not linear in  $\ket{v}$.

\vspace{2mm}
\noi
17.) It is easy to verify that the unitarity of the S-matrix, which is expressed by the equation $\int
\intd k\, S_{ik}(S_{jk})^* = \braket{v_i}{v_j}$, is not affected by the phase factors
 included in  (\ref{eq.Gin}) and (\ref {eq.Gout}).

\section{Examples}
\label{sec.example}

In this section we discuss some results that we obtained in certain specific models by means of
the application of (\ref{eq.Gin})-(\ref{eq.energy2}).
$H$ takes the form $H=H_K+gH_{\text{int}}$ in all cases, where $g$ is a coupling constant.
We studied five examples, which are the following.
The first one is the general case when
$H_K$ has a discrete nondegenerate and finite
spectrum. Although the Hamiltonian operators of this type do not describe scattering, from a technical point of view 
we found it interesting to consider this case.
The second example is the scattering of a single particle in short range spherical potentials in three dimensional space
in the framework of nonrelativistic quantum
mechanics. Scattering in a Dirac-delta
potential in one dimensional space is also briefly discussed.
The third example is
the scattering of a massive relativistic particle
on a defect in $1+1$ spacetime dimensions in the framework of quantum field theory. The particle is a real scalar
boson and the defect is localized at $x=0$.
The free particle Hamiltonian operator is
\begin{equation}
H_K=\frac{1}{2} \int_{-\infty}^\infty \intd x\, [:(\partial_t\Phi)^2 + (\partial_x\Phi)^2+m^2\Phi^2:];
\end{equation}
the interaction term is
\begin{equation}
H_{\text{int}}=:\Phi(0,0)^2:.
\end{equation}
The fourth example is the $\Phi^4$ theory in $3+1$ spacetime dimensions.
The free particle Hamiltonian operator is
\begin{equation}
H_K=\frac{1}{2} \int \intd^3 \mathbf{x}\, [:(\partial_t\Phi)^2 + (\partial_{\mathbf{x}}\Phi)^2+m^2\Phi^2:];
\end{equation}
the interaction term is
\begin{equation}
H_{\text{int}}=\int \intd^3 \mathbf{x}\, :\Phi^4:\ .
\end{equation}
The fifth example is the case when $H_K$ has a discrete, finite, but not necessarily nondegenerate spectrum. 

We took $H_K$ (i.e.\ the constant part
of $H$ as a linear function of $g$) as the 
reference Hamiltonian operator $H_0$ in all of the examples.

We used for our study the framework of the ordinary perturbation theory, as described in appendix \ref{sec.pert}.  In the first and third cases we did calculations up to second order in $g$; in the fourth example we also considered the third order. We considered all orders in the second example.
Feynman diagrams can be associated
with the various terms also in ordinary perturbation theory, if
it is applied to quantum field theories like the
$\Phi^4$ theory.

In the first example we found that our formalism
reproduces the results that can be obtained by Rayleigh-Schr\"odinger
perturbation theory (see e.g.\ \cite{Schwabl}) and gives the result that the S-matrix is the identity operator, in accordance with the remarks 
in 4.\ in section \ref{sec.disc}.

As the second example we discuss potential scattering. 
We expect that 
the modified formula (\ref{eq.smf}) and the standard formula (\ref{eq.qmsmat}) yield the same S-matrix in a large class of potential scattering problems.
In order to support this conjecture 
we will now discuss  the scattering of a single particle in short range spherical potentials in three dimensional space. We consider all orders of perturbation theory.

The basis vectors of the free particle Hamiltonian operator $H_K=-\frac{\partial^2}{\partial \mathbf{x}^2}$ are denoted by $\ket{\mathbf{k}}$; their wave function is
\beq
\ket{\mathbf{k}}=\frac{1}{(\sqrt{2\pi})^3} e^{i\mathbf{k}\mathbf{x}}
\eeq
The energy of $\ket{\mathbf{k}}$ is $\mathbf{k}^2$:
\beq
H_K\ket{\mathbf{k}}=\mathbf{k}^2\ket{\mathbf{k}}.
\eeq
The interaction Hamiltonian operator is 
\beq
H_{\text{int}}=V(\mathbf{x}),
\eeq
where $V(\mathbf{x})$ is the potential in which the scattering takes place. 
We assume that $V(\mathbf{x})$ is spherically symmetric, continuous and there exists a number $\rho>0$ such that it falls off at least as fast as $|\mathbf{x}|^{-3-\rho}$ at infinity (i.e.\ $|V(\mathbf{x})|<c|\mathbf{x}|^{-3-\rho}$ if $|\mathbf{x}|>r_0$, where $c$ and $r_0$ are suitable constants). 
Certain singularities at $x=0$ and at other points could be allowed as well (see chapter 2 of \cite{Taylor}).
The matrix elements of $H_{\text{int}}$ are (see (\ref{eq.notation}) for the notation on the left hand side)
\beq
\label{eq.mel}
\brakett{\mathbf{k}_1\mathbf{k}_2}=\frac{1}{(2\pi)^3}\int \intd^3 \mathbf{x}\   e^{i(\mathbf{k}_2-\mathbf{k}_1)\mathbf{x}}V(\mathbf{x})
=\frac{4\pi}{(2\pi)^3}\int_0^\infty r^2 \intd r\ \frac{\sin qr}{qr}V(r),
\eeq
where $r=|\mathbf{x}|$ and $q=|\mathbf{k}_1-\mathbf{k}_2|$.

In the following we will show that  $\lim_{T\to\infty}\Pi_{\mathbf{k},\epsilon}(-T)=1$ and 
$\lim_{T\to\infty}\Pi_{\mathbf{k},\epsilon}(T)=1$, which imply that  (\ref{eq.smf}) and (\ref{eq.qmsmat}) yield the same S-matrix.
We consider   $\lim_{T\to\infty}\Pi_{\mathbf{k},\epsilon}(-T)$ explicitly; the case of $\lim_{T\to\infty}\Pi_{\mathbf{k},\epsilon}(T)$ is very similar.
We use the notation $C_{\mathbf{k}_1\mathbf{k}_2,\kappa,-}$ for the coefficients in the Taylor series for  $\lim_{T\to
  \infty}\brakettt{\mathbf{k}_1}{U_\epsilon(0,-T)}{\mathbf{k}_2}$
 (see (\ref{eq.t1})). $\epsilon$ is kept fixed at some arbitrary positive value.

Under the above assumptions for $V(\mathbf{x})$ the matrix element $\brakett{\mathbf{k}_1\mathbf{k}_2}$ is finite for any $\mathbf{k_1}$ and $\mathbf{k_2}$ and $|\brakett{\mathbf{k}_1\mathbf{k}_2}|$ is also bounded.  As one can see from (\ref{eq.mel}), $|\brakett{\mathbf{k}_1\mathbf{k}_2}|$ falls off for  large $q$ at least as fast as $1/q$.
The coefficients $C_{\mathbf{k}_1\mathbf{k}_2,\kappa,-}$, $\kappa=1,2,\dots$, given by the integrals (\ref{eq.cijk1}), are also finite for any finite value of $\epsilon$ if one introduces a momentum cutoff. Power counting shows that the integrals giving these coefficients are finite without a cutoff as well, i.e.\ they are not ultraviolet divergent.

We introduce the regularized vectors (see appendix \ref{sec.reg}) as 
\beq
\ket{\mathbf{k},\mu}=
\int \intd^3 \mathbf{k}'\  \chi_\mu(k_1,k'_1) \chi_\mu(k_2,k'_2) \chi_\mu(k_3,k'_3) \ket{\mathbf{k}'}, 
\eeq
where $k_1$, $k_2$, $k_3$ are the components of $\mathbf{k}$, $\mu>0$ is a small real number  and $\chi_\mu(k,k')$ is the function given by
\begin{eqnarray}
\chi_\mu(k,k')= & \frac{1}{2k_0\mu} & \quad \mathrm{if}\ \ \ \ |k-k'|\le k_0\mu \\
\chi_\mu(k,k')= & 0 & \quad \mathrm{if}\ \ \ \ |k-k'|>k_0\mu\ , 
\end{eqnarray}
where $k_0>0$ is an arbitrary fixed constant. 
We use the notation $C_{\mathbf{k}_1\mathbf{k}_2,\kappa,\mu,-}$ for the coefficients 
in the series for  $\lim_{T\to
  \infty}\brakettt{\mathbf{k}_1,\mu}{U_\epsilon(0,-T)}{\mathbf{k}_2,\mu}$.
We have
\beq
C_{\mathbf{k}_1\mathbf{k}_2,\kappa,\mu,-}=\int \intd^3 \mathbf{k}_1'  \intd^3 \mathbf{k}_2'\ 
\chi_\mu(\mathbf{k}_1,\mathbf{k}_1')\chi_\mu(\mathbf{k}_2,\mathbf{k}_2') C_{\mathbf{k}_1'\mathbf{k}_2',\kappa,-}\ ,
\eeq
where $\chi_\mu(\mathbf{k},\mathbf{k}')=\chi_\mu(k_1,k'_1) \chi_\mu(k_2,k'_2) \chi_\mu(k_3,k'_3)$.
A coefficient $C_{\mathbf{k}_1\mathbf{k}_2,\kappa,\mu,-}$, where $\kappa>0$, has the limit $C_{\mathbf{k}_1\mathbf{k}_2,\kappa,-}$ as $\mu\to 0$, which is finite. From now on we will take $\mathbf{k}_1=\mathbf{k}_2=\mathbf{k}$, since this is the relevant case for $\Pi_{\mathbf{k},\epsilon}(-T)$. 
The zero order term in the Taylor series for  $\lim_{T\to
  \infty}\brakettt{\mathbf{k},\mu}{U_\epsilon(0,-T)}{\mathbf{k},\mu}$ is 
\beq
\braket{\mathbf{k},\mu}{\mathbf{k},\mu}=\frac{1}{(2k_0\mu)^3}, 
\eeq
which goes to infinity as $\mu\to 0$.
The Taylor series for $\lim_{T\to\infty}\Pi_{\mathbf{k},\epsilon,\mu}(-T)$ can be obtained from that for
 $\lim_{T\to
  \infty}\brakettt{\mathbf{k},\mu}{U_\epsilon(0,-T)}{\mathbf{k},\mu}$, taking into consideration (\ref{eq.phase1mod}).
The zero order term in the Taylor series for $\lim_{T\to\infty}\Pi_{\mathbf{k},\epsilon,\mu}(-T)$ is clearly
 $1$. The coefficients of the further terms in the series 
 will be polynomials (with zero constant term) of $C_{\mathbf{k}\mathbf{k},\kappa,\mu,-}/ \braket{\mathbf{k},\mu}{\mathbf{k},\mu}$ and 
$C_{\mathbf{k}\mathbf{k},\kappa,\mu,-}^*/ \braket{\mathbf{k},\mu}{\mathbf{k},\mu}$, where $\kappa$ can take positive integer values. 
However, $C_{\mathbf{k}\mathbf{k},\kappa,\mu,-}/ \braket{\mathbf{k},\mu}{\mathbf{k},\mu}$
and  $C_{\mathbf{k}\mathbf{k},\kappa,\mu,-}^*/ \braket{\mathbf{k},\mu}{\mathbf{k},\mu}$
 go to zero as $\mu\to 0$, since $C_{\mathbf{k}\mathbf{k},\kappa,\mu,-}$ has finite limit and $\braket{\mathbf{k},\mu}{\mathbf{k},\mu}$ goes to infinity. Thus all coefficients in the   Taylor series for $\lim_{T\to\infty}\Pi_{\mathbf{k},\epsilon,\mu}(-T)$ go to zero as $\mu\to 0$, only the zero order term remains, which is $1$. This completes our derivation of
$\lim_{T\to\infty}\Pi_{\mathbf{k},\epsilon}(-T)=1$.
In summary we can say that the essential point in the derivation is that $C_{\mathbf{k}\mathbf{k},\kappa,\mu,-}/\braket{\mathbf{k},\mu}{\mathbf{k},\mu}$, $\kappa=1,2,\dots$, go to zero as $\mu\to 0$. 

A similar argument as above can be applied to show that
the formulas (\ref{eq.energy1}) and (\ref{eq.energy2}) give zero for the corrections to
the energy eigenvalues. We discuss (\ref{eq.energy1}) only; the case of (\ref{eq.energy2}) is very similar.  
Using $H=H_K+gH_{\text{int}}$ we have 
\begin{eqnarray}
\lim_{T\to\infty}
\frac{\brakettt{\mathbf{k},\mu}{HU_\epsilon(0,-T)}{\mathbf{k},\mu}}{
\brakettt{\mathbf{k},\mu}{U_\epsilon(0,-T)}{\mathbf{k},\mu}} &  = & 
\lim_{T\to\infty}
\frac{\brakettt{\mathbf{k},\mu}{H_KU_\epsilon(0,-T)}{\mathbf{k},\mu}}{
\brakettt{\mathbf{k},\mu}{U_\epsilon(0,-T)}{\mathbf{k},\mu}} \nonumber  \\
& &
+\, g\lim_{T\to\infty}
\frac{\brakettt{\mathbf{k},\mu}{H_{\text{int}}U_\epsilon(0,-T)}{\mathbf{k},\mu}}{
\brakettt{\mathbf{k},\mu}{U_\epsilon(0,-T)}{\mathbf{k},\mu}}.
\end{eqnarray}
The limit of the first term on the right hand side as $\mu\to 0$ is the unperturbed energy $E_\mathbf{k}^0=\mathbf{k}^2$.
In the second term the numerator has a Taylor series in which all the coefficients have finite limit as $\mu\to 0$. The coefficients in the Taylor series for the denominator also have finite limit as $\mu\to 0$ with the exception of the zero order term $\braket{\mathbf{k},\mu}{\mathbf{k},\mu}$, which goes to infinity as $1/\mu^3$. All the coefficients of the Taylor series for the second term will therefore go to zero as $\mu\to 0$, thus the second term is zero, and 
$E_{\mathbf{k},in}=E_\mathbf{k}^0=\mathbf{k}^2$.

Very similar derivations to those presented above can be applied also, for instance, for the scattering of a particle in the Dirac-delta potential in one dimensional space. In this case the $H_K$ operator is the free particle Hamiltonian operator
$H_K=-\frac{\partial^2}{\partial x^2}$,
the interaction Hamiltonian operator is
$H_{\text{int}}=\delta(x)$.
We note that the S-matrix takes the form
\begin{equation}
\label{eq.1}
S_{k_1,k_2}=\delta(k_1-k_2)T(k_2) +\delta(k_1+k_2)R(k_2),
\end{equation}
and the exact
expressions for $T(k_2)$ and $R(k_2)$ can also be found; they are
\begin{equation}
\label{eq.2}
T(k_2)=\frac{i|k_2|}{i|k_2|+g/2},\qquad R(k_2)=\frac{g/2}{i|k_2|+g/2}.
\end{equation}
We also note that in perturbation theory there are infrared divergences in this model for \textit{in} and \textit{out} states with zero momentum.

Long range potentials, such as the Coulomb potential, present similar problems for the application of both 
 (\ref{eq.smf}) and the standard formula (\ref{eq.qmsmat}). In the case of the Coulomb potential, for instance, it is not immediately obvious what the matrix elements $\brakett{\mathbf{k}_1\mathbf{k}_2}$ are, since the integral on the right hand side of (\ref{eq.mel}) is not convergent for large $r$. A possible method of handling long range potentials is to introduce a shielding, i.e.\ to approximate them by short range potentials, as described in detail in \cite{Taylor}. This method is suitable for (\ref{eq.smf}) as well.

In the third example the interaction term breaks the Poincare symmetry of the free particle, so the 1-particle S-matrix $S_{k_1,k_2}$ is nontrivial. $S_{k_1,k_2}$ is given by the same expressions (\ref{eq.1}), (\ref{eq.2}) as the S-matrix for the scattering in the delta potential mentioned above, and our perturbative results were in agreement with this exact expression.
For the vacuum-vacuum S-matrix element we obtained $S_{\Omega_0,\Omega_0}=1+O(g^3)$, in agreement with 4.\ in section \ref{sec.disc}.
Concerning the energy levels, we found that all states, including the vacuum, get the same finite correction, the first order part of which is zero.
We considered particles in \textit{in} and \textit{out} states with nonzero momentum only, since at zero momentum infrared divergences occur.
We note that this model was also studied in a different approach in \cite{DMS}.

Turning to  the fourth example, we assumed, in accordance with the perturbative framework, that the $\Phi^4$ theory  describes an interacting massive relativistic boson.
It is well known that in this model one encounters ultraviolet divergent loop
 integrals in perturbation theory. Such divergences do not have significance for the present study and they do not cause much difficulty, therefore we were not concerned with them. One can introduce a momentum cutoff if one wants to have a regularization of these divergences.
 Radiative corrections on external lines appear first at third order, therefore we calculated S-matrix elements up to third order in this example. The results can be compared with those that can be obtained by means of the usual bare perturbation theory (in which $H$ is not renormalized). There are some differences between the presentations of standard perturbation theory in quantum field theory in various  textbooks; although these differences are not essential in principle, we note that we used \cite{PS} as reference.

For the vacuum-vacuum S-matrix element we obtained $S_{\Omega_0,\Omega_0}=1+O(g^4)$, in agreement with 4.\ in section \ref{sec.disc}. 
For the one-particle S-matrix we got $S_{\mathbf{k}_1,\mathbf{k}_2}=\delta^3(\mathbf{k}_1-\mathbf{k}_2)+O(g^4)$, again in agreement with the expectation stated in 4.\ in section \ref{sec.disc}.
As regards the two-particle S-matrix elements $S_{\mathbf{k}_3,\mathbf{k}_4;\mathbf{k}_1,\mathbf{k}_2}$, we obtained a result which is identical to the result that one can obtain in usual bare perturbation theory (see e.g.\ \cite{PS} for a description of bare perturbation theory). The result for $S_{\mathbf{k}_3,\mathbf{k}_4;\mathbf{k}_1,\mathbf{k}_2}$ is a formula that contains some loop 
integrals, which are ultraviolet divergent, of course.

\begin{figure}[t]
\begin{center}
\includegraphics[
  scale=0.3]{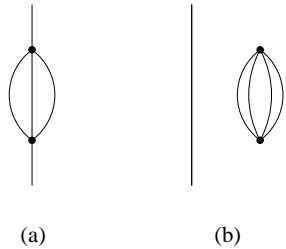}
\end{center}
\caption{\label{fig.diagA}
{\em Second order diagrams for the mass correction.}
}
\end{figure}

We also calculated the correction to the mass of the particle to second order. This correction can be obtained from the energy levels in the following way: The mass $m$ of a one-particle state is given by
$m=\sqrt{E^2-\mathbf{k}^2}$, where $E=E_{\mathbf{k}}-E_{\Omega_0}$ is its energy and $\mathbf{k}$ is its momentum. $E_{\mathbf{k}}$ and $E_{\Omega_0}$ denote the eigenvalue of the one-particle state and the eigenvalue of the vacuum state with respect to $H$.
We denote the expansion coefficients of $E$ and $m$ as follows:
\begin{equation}
E=E^0+gE^1+g^2E^2+...\ ,\qquad m=m_0+gm_1+g^2m_2+...\ .
\end{equation}
Similar notation as for $E$ applies to the expansion coefficients of $E_{\mathbf{k}}$ and $E_{\Omega_0}$ as well.
The series expansion of $m$ in terms of these coefficients up to second order in $g$ is
\begin{equation}
\label{eq.106}
m=m_0+g\frac{E^0E^1}{m_0}+g^2\left( \frac{E^0E^2}{m_0}-\frac{\mathbf{k}^2(E^1)^2}{2m_0^3}\right)+...\ ,
\end{equation}
where $m_0=\sqrt{(E^0)^2-\mathbf{k}^2}$. It is important in (\ref{eq.106}) that $m_0\ne 0$.
In the present example $E^1=0$, therefore (\ref{eq.106}) takes the form
$m=m_0+g^2 E^0E^2/m_0$. $E_{\mathbf{k}}^2$ and $E_{\Omega_0}^2$, and thus $E^2=E_{\mathbf{k}}^2-E_{\Omega_0}^2$, can be calculated using (\ref{eq.ein}).
The terms contributing to $E_{\mathbf{k}}^2$ are associated with the graphs shown in figure \ref{fig.diagA}.
The contributions corresponding to the graph \ref{fig.diagA}.b are equal to $E_{\Omega_0}^2$ and so are
cancelled out entirely by the
subtraction of $E_{\Omega_0}^2$. The terms contributing to $E^2$ will therefore be those corresponding
to the graph \ref{fig.diagA}.a. The result that we obtained for $m_2$ agrees with the result that can be obtained in bare perturbation theory in the standard formalism, in which the particle mass is given by the location of the pole of the propagator.
The result for the mass correction is a formula that contains an integral which is, of course, ultraviolet divergent.

Concerning multi-particle states, it is not difficult to verify that the additivity property mentioned in 5.) in section \ref{sec.disc} holds. Only those graphs give contributions to the energy of a multi-particle state which are totally disconnected, i.e.\ which do not describe any interaction between particles.
The contributions of those graphs  which describe some interaction other than self-interaction turn out to be zero in the $\mu\to 0$ limit (where $\mu$ is the regularization parameter for the eigenvectors of $H_0$, see appendix \ref{sec.reg}).

The fifth example is included in order to illustrate certain technical points related to 
the degeneracies of the eigenvalues of $H_0$.
Let us assume that $\ket{v_j}$ is a degenerate eigenvector.
Applying perturbation theory to calculate $\braket{v_i}{v_j,in}$, at first order we find that the $\epsilon\to 0^+$ limit is convergent only if $\brakettt{v_i}{H_{\text{int}}}{v_j}=0$ for all values of $i\ne j$ for which $E_i^0=E_j^0$. This condition is known from the degenerate Rayleigh-Schr\"odinger perturbation theory and can be satisfied by choosing the basis vectors within the eigenspace of $H_0$ containing $\ket{v_j}$ in such a way that the matrix $\brakettt{v_i}{H_{\text{int}}}{v_j}$ be diagonal within this eigenspace of $H_0$.
Assuming that this condition is satisfied, at second order we find that the $\epsilon\to 0^+$ limit exists if and only if 
\begin{equation}
\label{eq.cnd}
\sum_m \frac{\brakettt{v_i}{H_{\text{int}}}{v_m}\brakettt{v_m}{H_{\text{int}}}{v_j}}{E_m^0-E_j^0}=0
\end{equation}
holds for all values of $i\ne j$ for which $E_i^0=E_j^0$, where the summation has to be done over those values of $m$ for which $E_m^0\ne E_j^0$.
This condition is not exactly the same as the one in  Rayleigh-Schr\"odinger
perturbation theory; in the latter case, the equation (\ref{eq.cnd}) is required to hold only for those values of  $i\ne j$ for which  $E_i^0=E_j^0$ and  $E_i^1=E_j^1$ ($E_i^1$ and $E_j^1$ denote the coefficient of the first order corrections to $E_i^0$ and  $E_j^0$), i.e.\ for which $\ket{v_i,in}$ remains degenerate with $\ket{v_j,in}$ at first order. While the weaker condition of Rayleigh-Schr\"odinger perturbation theory can always be satisfied by choosing suitable basis vectors, the stronger condition that we have found cannot always be satisfied if the degeneracy of the eigenspace containing $\ket{v_j}$ is broken at first order. This problem can be handled by shifting those eigenvalues $E_i^0$ of $H_0$
for which $E_i^0=E_j^0$ but  $E_i^1\ne E_j^1$, i.e.\ by taking $H_0=H_K+\sum_i \delta E_i \ket{v_i}\bra{v_i}$, where $\delta E_i$ are the energy shifts. There is some freedom in the choice of  $\delta E_i$; one can take  $\delta E_i=g(E_i^1-E_j^1)$, for instance. $H_I=H-H_0$  also changes as a result of changing $H_0$; this should be taken into consideration in the perturbation theory. 
We expect that at higher orders we would find conditions similar to (\ref{eq.cnd}), which could also be handled in the same way as (\ref{eq.cnd}). 
In quantum field theory or in quantum mechanics one could also find conditions which are analogous to those described above. 
These conditions, which we do not discuss in detail, are often milder for those eigenvectors of $H_0$ which belong to continuous parts of the spectrum. If, for instance,   $\brakettt{v_i}{H_{\text{int}}}{v_j}=0$ were necessary in any case when  $i\ne j$ and $E_i^0=E_j^0$, then nontrivial scattering would be impossible at first order.
In some cases, for example when an interaction breaks the mass degeneracy of a multiplet, a suitable modification of an initial choice of basis vectors may be necessary; in some complicated quantum field theories a modification of $H_0$ may be necessary as well.
We can say that with our definition of the \textit{in} and \textit{out} states the divergences of the $\epsilon\to 0^+$ limit are related to certain kinds of breaking of degeneracies of the eigenvalues of $H_0$ by the interaction.

The calculations needed to obtain the above results are largely straightforward, nevertheless lengthy in some cases,
especially in the $\Phi^4$ theory.
The precise treatment of the $\epsilon\to 0^+$ limit is essential; in particular
the coefficients of $\epsilon$ in the denominators in (\ref{eq.cijk1}),
(\ref{eq.cijk2}), (\ref{eq.dijk})
must 
not be changed (see appendix \ref{sec.pert} for some further comments). In the second, third and fourth examples it is also important to keep in mind that the S-matrix elements and the components of the \textit{in} and \textit{out} states are distributions in general. One also encounters the problem mentioned in 14.\ in section \ref{sec.disc}. As we said there, one can introduce a regularization to handle this problem. We did this in the second example, otherwise it was satisfactory in our calculations to handle the Dirac-deltas in a formal manner.

In the first, third and fourth examples the $\brakettt{w}{U_\epsilon(T,-T)}{v}$, $\Pi_{w,\epsilon}(T)^*$ and $\Pi_{v,\epsilon}(-T)$ parts of (\ref{eq.smf}) are not individually convergent in the $\epsilon \to 0^+$ limit (after $T\to\infty$, of course). Up to the orders that we considered we found that these parts contain terms which diverge as $1/\epsilon$ in the third and fourth examples; in the first example terms diverging as $1/\epsilon^2$ also occur. In the third and fourth examples these divergent terms in $\brakettt{w}{U_\epsilon(T,-T)}{v}$ can be associated with diagrams having vacuum-vacuum diagram parts and radiative corrections on the external lines; the latter occur only in the fourth example. The phase factors $\Pi_{w,\epsilon}(T)^*$ and $\Pi_{v,\epsilon}(-T)$, however, give rise to further terms in the product $\Pi_{w,\epsilon}(T)^* \Pi_{v,\epsilon}(-T)
\brakettt{w}{U_\epsilon(T,-T)}{v}$, which turn out to have the effect that the divergences that would occur in the $\epsilon \to 0^+$ limit are eliminated;
all diagrams with vacuum-vacuum parts cancel out and the
radiative corrections on the external lines are replaced with a constant factor, which turns out to be equal to the square root of the field strength renormalization constant of the standard LSZ formalism.
This constant, which appears in the fourth example only, is given by a formula which contains ultraviolet divergent integrals.
The reader is invited to carry out these calculations
to see how the cancellations and the replacement take place.
In summary, we arrive effectively at the usual rules of perturbation theory, without the need to modify the choice $H_0=H_K$.
As we mentioned in the introduction and in section \ref{sec.rev}, if one wanted to apply  (\ref{eq.qmsmat}), then one would have to modify $H_0$ (assuming that $H$ is not changed) because of the vacuum-vacuum parts and the radiative corrections on the external lines.
We also verified in all of the examples up to second order that the $\epsilon\to 0^+$ limit of the components of the \textit{in} and \textit{out} states (i.e.\ the scalar products of the \textit{in} and \textit{out} states with the free states) is convergent up to second order (the ultraviolet divergences in the $\Phi^4$ theory are present, of course, for any values of $\epsilon$). The eigenvalue equations for the \textit{in} and \textit{out} states are also satisfied up to second order. Since the eigenvalue equation is an equation of vectors, we took the components of these vectors with respect to the free states and verified the equality of these components. As we mentioned earlier, it is important in the case of the second, third and fourth examples to keep in mind that one deals with distributions.

Finally, it is also worth emphasizing that  the results
$S_{\Omega_0,\Omega_0}=1$ and, in the $\Phi^4$ theory, $S_{\mathbf{k}_1,\mathbf{k}_2}=\delta^3(\mathbf{k}_1-\mathbf{k}_2)$ could be derived in our formalism in a straightforward way; it is not necessary to assume or to require that these equations hold.

\section{Conclusions}
\label{sec.concl}

In this paper we proposed a modification of the traditional general definition of the \textit{in} and \textit{out} states and of the S-matrix elements. The main components of our proposal are the following: phase factors are included in the definition of the \textit{in} and \textit{out} states in order to prevent oscillations occurring due to nonzero 
self-energies;
an adiabatic switching is used; and the requirement that an  \textit{in} or \textit{out} state has to have the same energy as the corresponding free state is dropped.
The divergences related to nonzero self-energies are considerably milder with the modified definition than 
with the usual one. With the modified definition divergences can occur due to certain breakings of degeneracies by the interaction. 

Using perturbation theory, we verified in certain quantum mechanical and quantum field
theoretical models that our definition yields the same S-matrix elements as the usual definitions.
In the quantum field theoretical examples we did the verification up to second and third orders.
For those cases of potential scattering that we discussed we presented a derivation of this result to all orders.
In the quantum field theoretical examples it was not necessary to do any renormalization to remove the
divergences related to nonzero vacuum self-energies and self-masses.
We also calculated energy and mass corrections and obtained results which are in agreement with expectations and
with the results that can be obtained using standard definitions.
We expect that these results extend to arbitrary orders of perturbation theory, and similar results hold for other models in quantum field theory as well.

\section*{Acknowledgments}

I would like to thank K\'alm\'an T\'oth for carefully reading the manuscript and for several useful suggestions. I acknowledge partial support by the Italian grant 2007JHLPEZ.

\appendix

\section{Perturbation theory}
\label{sec.pert}

In this appendix we describe ordinary perturbation theory, which is our tool in
section \ref{sec.example}. We include this part in the interest of completeness and
because the coefficients of $\epsilon$ in the denominators in
(\ref{eq.cij1m}), (\ref{eq.cijk1}),
(\ref{eq.cij1p}), (\ref{eq.cijk2}), (\ref{eq.dij1A}), (\ref{eq.dijk}), (\ref{eq.TT1}),
(\ref{eq.TT2}) and (\ref{eq.TT3})
differ from those that can be found in similar formulas in a large part of the literature. However, these coefficients should generally be treated precisely in the calculations; otherwise one obtains wrong results in some cases. The coefficients of $\epsilon$ are characteristic of 
the adiabatic switching prescription and of the exponential switching function.

We assume that $H$ takes the form
\begin{equation}
H=H_0+gH_{\text{int}},
\end{equation}
where $g$ is a coupling constant and $H_0$ is the reference  Hamiltonian operator.
The $H_I$ operator is thus $H_I=gH_{\text{int}}$.
We also assume that an orthonormal basis $\{\ket{v_i}\}$ of eigenvectors of $H_0$ is given.
The perturbation series which we shall deal with will be Taylor series in $g$.
We assume that $H_0$, $H_{\text{int}}$ and $\{\ket{v_i}\}$ is independent of
$g$. If these quantities depend on $g$, then the series written below will not
be Taylor series in $g$; in order to obtain the power series one should  expand the various terms of these series into powers of $g$ and collect the terms proportional to the same power of $g$.

A series expansion for
$\brakettt{v_i}{U_\epsilon(t_2,t_1)}{v_j}$ can be obtained in the following way. One expands
$U_\epsilon(t_2,t_1)$ into the Dyson series
\begin{eqnarray}
&&U_\epsilon(t_2,t_1)  =T\exp\left[-\frac{i}{\hbar}\int_{t_1}^{t_2}gH_{\text{int},\epsilon}(t)\,\intd
    t\right]\nonumber\\
&&=I+\sum_{N=1}^{\infty}\left(\frac{-i}{\hbar}\right)^N\frac{g^N}{N!}\int_{t_1}^{t_2}\intd\tau_1\intd\tau_2\dots
\intd\tau_N\,
T[H_{\text{int},\epsilon}(\tau_1)H_{\text{int},\epsilon}(\tau_2)\dots
    H_{\text{int},\epsilon}(\tau_N)]\nonumber\\
&& =I+\sum_{N=1}^\infty g^N\left(\frac{-i}{\hbar}\right)^N \int_{t_1}^{t_2}\intd\tau_1 \int_{t_1}^{\tau_1}\intd\tau_2
\int_{t_1}^{\tau_2}\intd\tau_3 \dots \nonumber\\
&& \hspace{5cm} \dots\int_{t_1}^{\tau_{N-1}}\intd\tau_N\,
H_{\text{int},\epsilon}(\tau_1)H_{\text{int},\epsilon}(\tau_2)\dots H_{\text{int},\epsilon}(\tau_N),
\label{eq.ser}
\end{eqnarray}
where $H_{\text{int},\epsilon}(t)=e^{-\epsilon |t|} e^{\frac{i}{\hbar}H_0t}H_{\text{int}} e^{-\frac{i}{\hbar}H_0t}$.
One inserts unit operators in the form
\begin{equation}
I=\int \intd m\, \ket{v_m}\bra{v_m}
\end{equation}
between
the $H_{\text{int},\epsilon}(\tau)$-s in (\ref{eq.ser}):
\begin{eqnarray}
&&\brakettt{v_i}{H_{\text{int},\epsilon}(\tau_1)H_{\text{int},\epsilon}(\tau_2)\dots
  H_{\text{int},\epsilon}(\tau_N)}{v_j}
=\nonumber\\
&&\int  {\intd m_1 \intd m_2 \dots \intd m_{N-1}}\,
\brakettt{v_i}{H_{\text{int},\epsilon}(\tau_1)}{v_{m_1}}
\brakettt{v_{m_1}}{H_{\text{int},\epsilon}(\tau_2)}{v_{m_2}}
\dots
\brakettt{v_{m_{N-1}}}{H_{\text{int},\epsilon}(\tau_N)}{v_{j}},\nonumber\\
\end{eqnarray}
and then performs the integrals over
the $\tau$-s, which  is possible since
\begin{equation}
\label{eq.expon}
\brakettt{v_n}{H_{\text{int},\epsilon}(\tau)}{v_m}=
\brakettt{v_n}{H_{\text{int}}}{v_m}\exp\left[\frac{i}{\hbar}(E_n^0-E_m^0)\tau-\epsilon
|\tau |
\right]
\end{equation}
is a simple exponential function. After the  integrations over
the $\tau$-s are performed,
the integrations $\int  {\intd m_1 \intd m_2 \dots \intd m_{N-1}}$ over the intermediate states still remain in the formula.

Taylor series that we obtained for
$\lim_{T\to\infty}\brakettt{v_i}{U_\epsilon(0,-T)}{v_j}$,
$\lim_{T\to\infty}\brakettt{v_i}{U_\epsilon(T,0)}{v_j}$
and
$\lim_{T\to\infty}\brakettt{v_i}{U_\epsilon(T,-T)}{v_j}$
are presented below.
These formulas arise after performing the integrations over
the $\tau$-s.

The series written below  for
 $\brakettt{v_i}{U_\epsilon(t_1,t_2)}{v_j}$,
$\lim_{T\to\infty}\brakettt{v_i}{U_\epsilon(0,-T)}{v_j}$,\\
$\lim_{T\to\infty}\brakettt{v_i}{U_\epsilon(T,0)}{v_j}$
and
$\lim_{T\to\infty}\brakettt{v_i}{U_\epsilon(T,-T)}{v_j}$
can  be used to obtain the power
series for the various quotients and products in section \ref{sec.main}.
We list the coefficients of the series that we obtained for the \textit{in} and
 \textit{out} states, for the S-matrix, and for the eigenvalues of $H$ up to second
order.
 Coefficients for some
further quantities are also given.

We remark that the matrix elements
$\brakettt{v_i,in}{A}{v_j,in}$, $\brakettt{v_i,out}{A}{v_j,out}$,\\
$\brakettt{v_i,in}{A}{v_j,out}$ and $\brakettt{v_i,out}{A}{v_j,in}$
of any  operator $A$ between \textit{in} and
\textit{out} states can also be calculated perturbatively if the matrix elements
$\brakettt{v_i}{A}{v_j}$ of $A$ between the eigenstates of $H_0$ are known.
For instance,
\begin{equation}
\brakettt{v_i,in}{A}{v_j,in}=
\int \intd m_1 \intd m_2\,
\braket{v_i,in}{v_{m_1}}
\brakettt{v_{m_1}}{A}{v_{m_2}}
\braket{v_{m_2}}{v_j,in},
\end{equation}
and here the perturbative results for $\braket{v_i,in}{v_{m_1}}$ and
$\braket{v_{m_2}}{v_j,in}$ can be used to obtain a power series for $\brakettt{v_i,in}{A}{v_j,in}$.

We begin our list of formulas with formulas for $\lim_{T\to\infty}\brakettt{v_i}{U_\epsilon(0,-T)}{v_j}$,\\
$\lim_{T\to\infty}\brakettt{v_i}{U_\epsilon(T,0)}{v_j}$
and
$\lim_{T\to\infty}\brakettt{v_i}{U_\epsilon(T,-T)}{v_j}$.
We use the notation
\begin{equation}
\label{eq.notation}
P(ab)=\frac{i}{\hbar}(E_a^0-E_b^0),\qquad
\brakett{ab}=\brakettt{a}{H_{\text{int}}}{b}.
\end{equation}

\noindent
A.) For  $\lim_{T\to
  \infty}\brakettt{v_i}{U_\epsilon(0,-T)}{v_j}$  we obtain
\begin{equation} 
\label{eq.t1}
\lim_{T\to
  \infty}\brakettt{v_i}{U_\epsilon(0,-T)}{v_j}=
\sum_{k=0}^\infty \left(\frac{-i}{\hbar}\right)^k g^k C_{ij,k,-}\ ,
\end{equation}
where
\begin{equation}
\label{eq.cij1m}
C_{ij,0,-}=\braket{v_i}{v_j},\qquad
C_{ij,1,-}=\frac{\brakett{ij}}{P(ij)+\epsilon},
\end{equation}
\begin{eqnarray}
&&C_{ij,k,-}=\nonumber\\
&&\int {\intd m_1 \intd m_2 \dots \intd m_{k-1}} \frac{
  \brakett{im_{k-1}}\brakett{m_{k-1}m_{k-2}}\dots
   \brakett{m_2m_1}
   \brakett{m_1j}}{[P(ij)+k\epsilon][P(m_{k-1}j)+(k-1)\epsilon]\dots
  [P(m_{2}j)+2\epsilon][P(m_{1}j)+\epsilon]}.\nonumber\\
\label{eq.cijk1}
\end{eqnarray}

\noindent
B.) For  $\lim_{T\to
  \infty}\brakettt{v_i}{U_\epsilon(T,0)}{v_j}$ we obtain
\begin{equation}
\label{eq.t2}
\lim_{T\to
  \infty}\brakettt{v_i}{U_\epsilon(T,0)}{v_j}=
\sum_{k=0}^\infty \left(\frac{-i}{\hbar}\right)^k g^k C_{ij,k,+}\ ,
\end{equation}
where
\begin{equation}
\label{eq.cij1p}
C_{ij,0,+}=\braket{v_i}{v_j},\qquad
C_{ij,1,+}=\frac{\brakett{ij}}{P(ji)+\epsilon},
\end{equation}
\begin{eqnarray}
&&C_{ij,k,+}=\nonumber\\
&&\int {\intd m_1 \intd m_2 \dots \intd m_{k-1}} \frac{
  \brakett{im_{1}}\brakett{m_{1}m_{2}}\dots
   \brakett{m_{k-2}m_{k-1}} \brakett{m_{k-1}j}}{[P(m_1
  i)+\epsilon][P(m_{2}i)+2\epsilon]\dots
[P(m_{k-1}i)+(k-1)\epsilon][P(ji)+k\epsilon]}.\nonumber\\
\label{eq.cijk2}
\end{eqnarray}

\noindent
C.) For  $\lim_{T\to
  \infty}\brakettt{v_i}{U_\epsilon(T,-T)}{v_j}$ we obtain
\begin{equation} 
\lim_{T\to
  \infty}\brakettt{v_i}{U_\epsilon(T,-T)}{v_j}=
\sum_{k=0}^\infty \left(\frac{-i}{\hbar}\right)^k g^k D_{ij,k}\ ,
\end{equation}
where
\begin{equation}
\label{eq.dij1A}
D_{ij,0}=\braket{v_i}{v_j},\qquad
D_{ij,1}=\left[\frac{1}{P(ij)+\epsilon}+\frac{1}{P(ji)+\epsilon}\right]\brakett{ij},
\end{equation}
\begin{equation}
\label{eq.dijk0}
D_{ij,k}=
\int {\intd m_1 \intd m_2 \dots \intd m_{k-1}} \,
\brakett{i m_{1}}\brakett{m_{1}m_{2}}\dots
   \brakett{m_{k-2}m_{k-1}} \brakett{m_{k-1}j}\,
 \tilde{D}_{ij,k}\ ,
\end{equation}
\begin{equation}
\label{eq.dijk}
\tilde{D}_{ij,k}=
\sum_{l=0}^k
\left[
\left(\prod_{r=1}^l   \frac{1}{P(m_r i)+r\epsilon}\right) \times
\left(\prod_{s=l}^{k-1}   \frac{1}{P(m_s j)+(k-s)\epsilon}\right)
\right],
\end{equation}
\begin{equation}
m_0= i,\qquad m_k = j.
\end{equation}
Up to $k=2$ we have
\begin{eqnarray}
&&\hspace{-1cm}\lim_{T\to
  \infty}\brakettt{v_i}{U_\epsilon(T,-T)}{v_j} =
\braket{v_i}{v_j}+
\frac{-i}{h}g\left[\frac{1}{P(ij)+\epsilon}+\frac{1}{P(ji)+\epsilon}\right]\brakett{ij} \nonumber\\
&&\hspace{-1cm} + \left(\frac{-i}{h}\right)^2g^2 \int \intd m\,
\brakett{im}\brakett{mj} \nonumber\\
&&\times \left[\frac{1}{P(mi)+\epsilon}\frac{1}{P(ji)+2\epsilon}+\frac{1}{P(mj)+\epsilon}\frac{1}{P(ij)+2\epsilon}+\frac{1}{P(mj)+\epsilon}\frac{1}{P(mi)+\epsilon}\right].
\label{eq.AA}
\end{eqnarray}
The formulas in A.) and B.) and C.) can also be written as
\begin{eqnarray}
&&\hspace{-1cm}
\lim_{T\to
  \infty}{U_\epsilon(0,-T)}\ket{v_j}=\nonumber\\
&&\left[\rule{0cm}{0.7cm}  I
+\sum_{k=1}^\infty \left(\frac{-i}{\hbar}\right)^k g^k
\frac{1}{\frac{i}{\hbar}(H_0-E_j^0)+k\epsilon}H_{\text{int}}
\frac{1}{\frac{i}{\hbar}(H_0-E_j^0)+(k-1)\epsilon}H_{\text{int}}\ \dots \right.\nonumber\\
&&\left. \dots \ 
\frac{1}{\frac{i}{\hbar}(H_0-E_j^0)+2\epsilon}H_{\text{int}}
\frac{1}{\frac{i}{\hbar}(H_0-E_j^0)+\epsilon}H_{\text{int}}\rule{0cm}{0.7cm}\right] \ket{v_j}
\label{eq.TT1}
\end{eqnarray}
\begin{eqnarray}
&&\hspace{-1cm}\lim_{T\to
  \infty} \bra{v_i}{U_\epsilon(T,0)}=\nonumber\\
&&\bra{v_i} \left[\rule{0cm}{0.7cm}
I+
\sum_{k=1}^\infty \left(\frac{-i}{\hbar}\right)^k g^k
H_{\text{int}}\frac{1}{\frac{i}{\hbar}(H_0-E_i^0)+\epsilon}H_{\text{int}}
\frac{1}{\frac{i}{\hbar}(H_0-E_i^0)+2\epsilon}\ \dots \right. \nonumber\\
&&\dots \ \left.
H_{\text{int}}\frac{1}{\frac{i}{\hbar}(H_0-E_i^0)+(k-1)\epsilon}
H_{\text{int}}\frac{1}{\frac{i}{\hbar}(H_0-E_i^0)+k\epsilon}
\rule{0cm}{0.7cm} \right]
\label{eq.TT2}
\end{eqnarray}
\begin{eqnarray}
&&\hspace{-1cm}\lim_{T\to
  \infty} \brakettt{v_i}{U_\epsilon(T,-T)}{v_j}=\nonumber\\
&&\bra{v_i} \left[\rule{0cm}{0.7cm}
I+
\sum_{k=1}^\infty \sum_{l=0}^k \left(\frac{-i}{\hbar}\right)^k g^k
H_{\text{int}}\frac{1}{\frac{i}{\hbar}(H_0-E_i^0)+\epsilon} \dots H_{\text{int}}
\frac{1}{\frac{i}{\hbar}(H_0-E_i^0)+l\epsilon}\ \times \right. \nonumber\\
&&\times \left.
\frac{1}{\frac{i}{\hbar}(H_0-E_j^0)+(k-l)\epsilon}H_{\text{int}}\dots
\frac{1}{\frac{i}{\hbar}(H_0-E_j^0)+\epsilon}H_{\text{int}}
\rule{0cm}{0.7cm} \right]\ket{v_j}.
\label{eq.TT3}
\end{eqnarray}

Below we present perturbation series up to second order 
for the components of the \textit{in} and  \textit{out} states (\ref{eq.Gin}) and (\ref{eq.Gout}), for the
S-matrix elements  (\ref{eq.smf}), for the energy eigenvalues
(\ref{eq.energy1}) and (\ref{eq.energy2}), for the phase
factor (\ref{eq.phase1}), and for the normalization constant (\ref{eq.x1}), using the formulas
written above. 
$\braket{v_i}{v_j}$ are treated formally in the calculations (see 14.\ in section \ref{sec.disc}). $\delta_{ij}$ denotes
$\braket{v_i}{v_j}/\braket{v_i}{v_i}$.\\

\noindent
D.) For the \textit{in} state (\ref{eq.Gin})
we found 
\begin{equation}
 \lim_{T\to\infty} \Pi_{j,\epsilon}(-T)
\brakettt{v_i}{U_\epsilon(0,-T)}{v_j}=
\braket{v_i}{v_j}+g\left(\frac{-i}{\hbar}\right){c}_{ij,1,-}+g^2\left(\frac{-i}{\hbar}\right)^2  {c}_{ij,2,-}+O(g^3),
\end{equation}
where
\begin{equation}
{c}_{ij,1,-}=
\frac{\brakett{ij} (1-\delta_{ij})}{P(ij)+\epsilon},
\end{equation}
\begin{eqnarray}
{c}_{ij,2,-} & = &\int \intd m
\frac{\brakett{im}\brakett{mj}}{[P(ij)+2\epsilon][P(mj)+\epsilon]}
+
\int \intd m
\frac{\brakett{jm}\brakett{mj}\delta_{ij}}{4\epsilon[\epsilon-P(mj)]}
-
\int \intd m
\frac{\brakett{jm}\brakett{mj}\delta_{ij}}{4\epsilon[\epsilon+P(mj)]}\nonumber\\
&&+
\frac{\brakett{jj}^2\delta_{ij}}{\braket{v_j}{v_j} 2\epsilon^2}
-
\frac{\brakett{ij}\brakett{jj}}{\braket{v_j}{v_j} \epsilon [P(ij)+\epsilon]}.
\end{eqnarray}

\noindent
E.)  For the \textit{out} state (\ref{eq.Gout})
we found
\begin{equation}
\lim_{T\to\infty}  \Pi_{i,\epsilon}(T)^*
\brakettt{v_i}{U_\epsilon(T,0)}{v_j}=
\braket{v_i}{v_j}+g\left(\frac{-i}{\hbar}\right){c}_{ij,1,+}+g^2\left(\frac{-i}{\hbar}\right)^2  {c}_{ij,2,+}+O(g^3),
\end{equation}
where
\begin{equation}
{c}_{ij,1,+}=
\frac{\brakett{ij} (1-\delta_{ij})}{P(ji)+\epsilon},
\end{equation}
\begin{eqnarray}
{c}_{ij,2,+} & = & \int \intd m
\frac{\brakett{im}\brakett{mj}}{[P(ji)+2\epsilon][P(mi)+\epsilon]}
+
\int \intd m
\frac{\brakett{im}\brakett{mi}\delta_{ij}}{4\epsilon[\epsilon-P(mi)]}
-
\int \intd m
\frac{\brakett{im}\brakett{mi}\delta_{ij}}{4\epsilon[\epsilon+P(mi)]}\nonumber\\
&&+
\frac{\brakett{ii}^2\delta_{ij}}{\braket{v_i}{v_i} 2\epsilon^2}
-
\frac{\brakett{ii}\brakett{ij}}{\braket{v_i}{v_i} \epsilon [P(ji)+\epsilon]}.
\end{eqnarray}

\noindent
F.) For the S-matrix elements (\ref{eq.smf}) we obtained
\begin{eqnarray}
&&\lim_{T\to\infty}
 \Pi_{i,\epsilon}(T)^*  \Pi_{j,\epsilon}(-T)  \brakettt{v_i}{U_\epsilon(T,-T)}{v_j} = \nonumber \\
&&=\braket{v_i}{v_j}
+g\left(\frac{-i}{\hbar}\right)d_{ij,1}
+g^2\left(\frac{-i}{\hbar}\right)^2  d_{ij,2}+O(g^3),
\label{eq.dij}
\end{eqnarray}
where
\begin{equation}
\label{eq.dij1}
d_{ij,1}=\brakett{ij} (1-\delta_{ij})\left[\frac{1}{P(ij)+\epsilon}
+\frac{1}{P(ji)+\epsilon}\right],
\end{equation}
\begin{eqnarray}
d_{ij,2} & = & \int \intd m\,
\brakett{im}\brakett{mj}\nonumber \\
&&\times
\left[  \frac{1}{P(ij)+2\epsilon}\frac{1}{P(mj)+\epsilon}+
\frac{1}{P(ji)+2\epsilon} \frac{1}{P(mi)+\epsilon}
+ \frac{1-\delta_{mj}}{P(mj)+\epsilon}\frac{1-\delta_{mi}}{P(mi)+\epsilon}
 \right]\nonumber\\
&&+
\int \intd m
\frac{\brakett{im}\brakett{mi}\delta_{ij}}{2\epsilon[\epsilon-P(mi)]}
-
\int \intd m
\frac{\brakett{im}\brakett{mi}\delta_{ij}}{2\epsilon[\epsilon+P(mi)]}
+
\frac{\brakett{ii}^2\delta_{ij}}{\braket{v_i}{v_i} 2\epsilon^2}
+
\frac{\brakett{jj}^2\delta_{ij}}{\braket{v_j}{v_j} 2\epsilon^2}\nonumber
\\
&&-
\frac{\brakett{ii}\brakett{ij}}{\braket{v_i}{v_i} \epsilon [P(ji)+\epsilon]}
-
\frac{\brakett{ij}\brakett{jj}}{\braket{v_j}{v_j} \epsilon [P(ij)+\epsilon]}.
\label{eq.dij2}
\end{eqnarray}

\noindent
G.) We obtained the following formula for the eigenvalues of the \textit{in} states:
\begin{eqnarray}
E_{i,in,\epsilon}(g) & = &
\frac{\lim_{T\to\infty}
  \brakettt{v_i}{HU_\epsilon(0,-T)}{v_i}}{\lim_{T\to\infty}
  \brakettt{v_i}{U_\epsilon(0,-T)}{v_i}} \nonumber\\
& = & E_i^0 + g
\frac{\brakett{ii}}{\braket{v_i}{v_i}}
+g^2\frac{-i}{\hbar}
\int \intd m
\frac{\brakett{im}\brakett{mi}(1-\delta_{im})}{\braket{v_i}{v_i}[P(mi)+\epsilon]}
+O(g^3).
\label{eq.ein}
\end{eqnarray}
For the eigenvalues of the \textit{out} states we obtained
\begin{eqnarray}
E_{i,out,\epsilon}(g) & = &
\frac{\lim_{T\to\infty}
  \brakettt{v_i}{HU_\epsilon(0,-T)}{v_i}}{\lim_{T\to\infty}
  \brakettt{v_i}{U_\epsilon(0,-T)}{v_i}} \nonumber\\
 & = &E_i^0 + g
\frac{\brakett{ii}}{\braket{v_i}{v_i}}
+g^2\frac{-i}{\hbar}
\int \intd m
\frac{\brakett{im}\brakett{mi}(1-\delta_{im})}{\braket{v_i}{v_i}[P(mi)-\epsilon]}
+O(g^3).
\label{eq.eout}
\end{eqnarray}

\noindent
H.) For the phase
factor (\ref{eq.phase1}) we obtained
\begin{eqnarray}
&&\lim_{T\to\infty} \Pi_{i,\epsilon}(-T)  =1+g\frac{i}{\hbar}\frac{\brakett{ii}}{\braket{v_i}{v_i}\epsilon}   \nonumber\\
&&+g^2\frac{1}{\hbar^2}\left[
-\frac{1}{2\epsilon^2}\frac{\brakett{ii}^2}{\braket{v_i}{v_i}^2}-
\int \intd m
\frac{\brakett{im}\brakett{mi}}{4\epsilon[\epsilon-P(mi)]\braket{v_i}{v_i}}+
\int \intd m
\frac{\brakett{im}\brakett{mi}}{4\epsilon[\epsilon+P(mi)]\braket{v_i}{v_i}}\right]\nonumber \\
&&+O(g^3).
\label{eq.pi}
\end{eqnarray}
We found that
$\lim_{T\to\infty} \Pi_{i,\epsilon}(-T) = \lim_{T\to\infty} \Pi_{i,\epsilon}(T)^* $ up to second order.\\

\noindent
I.)  For the normalization constant (\ref{eq.x1}) we obtained
\begin{eqnarray}
&&\lim_{T\to\infty}\sqrt{X_{i,\epsilon}(-T)}=\braket{v_i}{v_i} \nonumber\\
&&+g^2\left(\frac{-i}{\hbar}\right)^2 \left[
-\frac{\brakett{ii}^2}{2\braket{v_i}{v_i}\epsilon^2}
+\int \intd m \frac{\brakett{im}\brakett{mi}}{4\epsilon [\epsilon-P(mi)]}
+\int \intd m \frac{\brakett{im}\brakett{mi}}{4\epsilon [\epsilon+P(mi)]}
\right]+
O(g^3),\nonumber\\
\end{eqnarray}
and 
$\lim_{T\to\infty}\sqrt{X_{i,\epsilon}(-T)}=\lim_{T\to\infty}\sqrt{X_{i,\epsilon}(T)}$
up to second order.

\section{Regularization of the eigenvectors}
\label{sec.reg}

It was mentioned in section \ref{sec.disc} that complications may arise
when one calculates  (\ref{eq.smf})  using perturbation theory or other methods, due to the fact that often
$\braket{v_i}{v_i}$ is not a finite number and
 $\braket{v_i}{v_j}$ has to be regarded as a distribution.
One way to tackle this situation
is to introduce a regularization of the eigenvectors  $\ket{v_i}$.
In this appendix we outline such a regularization method.

One takes
superpositions
\begin{equation}
\ket{v_i,\mu}=\int \intd a\ f_\mu (i,a)\ket{v_a},
\end{equation}
where $\mu$ is a regularization parameter and  $f_\mu (i,a)$ are suitable
 functions of $i$ and $a$ with the property
that
\begin{equation}
\braket{v_i,\mu}{v_j,\mu}=  \int \intd a\
f_{\mu}(i,a)^* f_{\mu}(j,a)
\end{equation}
is finite for all $i$ and $j$. In the limit $\mu \to 0$ the original
 $\ket{v_i}$ vectors should be recovered. In particular,
$\lim_{\mu\to 0}\braket{v_i,\mu}{v_j,\mu}=\delta(i,j)$ should hold, where $\delta(i,j)$ is
the Dirac-delta distribution.

In the various formulas, in particular in  (\ref{eq.Gin})-(\ref{eq.energy2}),
one should replace $\ket{v_i}$ and $\ket{v_j}$ by
 $\ket{v_i,\mu}$ and  $\ket{v_j,\mu}$. After this replacement the calculations can be carried out, keeping $\mu$ and $\epsilon$ finite.
Then one should take the $\mu \to
0$ limit and, subsequently, the limit
$\epsilon\to 0^+$.

In the case of the S-matrix elements, for example, one has
\begin{equation}
\label{eq.smfmod}
S_{ij}=
 \lim_{\epsilon\to 0^+} \lim_{\mu\to 0} \lim_{T\to\infty}
 \Pi_{i,\epsilon,\mu}(T)^*    \Pi_{j,\epsilon,\mu}(-T)
\brakettt{v_i,\mu}{U_\epsilon(T,-T)}{v_j,\mu},
\end{equation}
where
\begin{equation}
\label{eq.phase1mod}
 \Pi_{i,\epsilon,\mu}(T) =
\frac{\sqrt{\brakettt{v_i,\mu}{U_\epsilon(T,0)}{v_i,\mu}\brakettt{v_i,\mu}{U_\epsilon(0,T)}{v_i,\mu}}}{
\brakettt{v_i,\mu}{U_\epsilon(0,T)}{v_i,\mu}}.
\end{equation}

\end{document}